# USE OF A BIOMECHANICAL TONGUE MODEL TO PREDICT THE IMPACT OF TONGUE SURGERY ON SPEECH PRODUCTION

Stéphanie Buchaillard[1,3], Muriel Brix[2], Pascal Perrier[1], Yohan Payan[3]

[1]ICP /GIPSA-lab, UMR CNRS 5216, INP Grenoble, France
[2]University Hospital, Grenoble, France
[3]TIMC-IMAG, UMR CNRS 5525, Université Joseph Fourier, Grenoble, France

**Abstract:** This paper presents predictions of the consequences of tongue surgery on speech production. For this purpose, a 3D finite element model of the tongue is used that represents this articulator as a deformable structure in which tongue muscles anatomy is realistically described. Two examples of tongue surgery, which are common in the treatment of cancers of the oral cavity, are modelled, namely a hemiglossectomy and a large resection of the mouth floor. In both cases, three kinds of possible reconstruction are simulated, assuming flaps with different stiffness. Predictions are computed for the cardinal vowels /i, a, u/ in the absence of any compensatory strategy, i.e. with the same motor commands as the one associated with the production of these vowels in non-pathological conditions. The estimated vocal tract area functions and the corresponding formants are compared to the ones obtained under normal conditions.

## I. INTRODUCTION

Resection surgery can be required in case of a cancerous tongue tumour or for particular pathologies like a macroglossia, characterized by an abnormally voluminous tongue. In case of noticeable loss of bulk or volume, the tongue is reconstructed using a local or distant flap in order to limit the functional consequences, of which choice is still a debated question.

The surgical procedure can impair the tongue mobility and tongue deformation capabilities, which can deteriorate the three basic functions of the human life, namely mastication, swallowing and speech. The surgery consequences can then induce a noticeable decrease of the patients's quality of life. The current project aims at developing some software that would allow surgeons to predict the consequences of a tongue resection for a given patient, using a 3D biomechanical model of the oral cavity, combined with a synthesizer based on the vocal tract area function. By now, the model has been tested for two common exeresis schemes for a particular subject. In this paper, we first introduce briefly the model used for this study and the implementation followed for two glossectomies (resection and reconstruction). Then we present the results obtained for the cardinal vowels /i, a, u/ in terms of formants deviations and tongue mobility, compared to the non pathological case.

## II. METHODS

**A 3D biomechanical Tongue model**
The 3D biomechanical model of the oral cavity used in this study was originally designed by Gérard *et al.* [1] and was further enhanced for speech production control [2] (Fig. 1). The tongue and the hyoid bone are represented by mobile 3D volumetric meshes, while the jaw, teeth, palate, and pharynx are modelled by static surface elements describing the oral cavity limits with which the tongue interacts due to mechanical contacts.

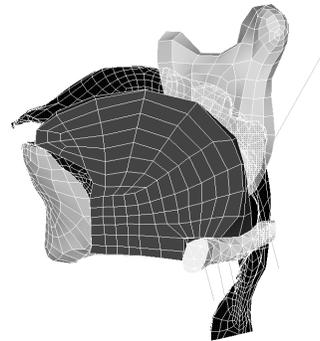

**Figure 1:** 3D model of the tongue in the midsagittal plane (apex on the left).

**Modelling tongue resections**
To model a surgical resection followed by a flap reconstruction, the muscles fibres located in the resected area are removed and the biomechanical properties of the corresponding elements are modified to account for the elastic properties of the flap. Tissues stiffness identical to the one of the passive tissues, 5 times smaller or 6 times higher are considered. In addition, since little is known about the force generation capabilities of muscles that have been partially shortened, three options were tested for the activation of sectioned fibres: 1) no activation, 2) low activation or 3) similar level of activation as in the normal case. Additional details about our general modelling approach can be found in [3].

The first simulated surgery corresponds to a left hemiglossectomy (Fig. 2, right panel). The left part of the styloglossus is removed as well as the left anterior parts of the longitudinal muscle, of the transversalis, and of the verticalis, and the upper part of the left hyoglossus. The medium and anterior parts of the left genioglossus are nearly entirely removed, whereas its posterior part is only partially affected.

The second simulated surgery corresponds to a large mouth floor resection (Fig. 2, left panel). In that case, the mobile tongue is totally preserved. The anterior part of the genioglossus is removed as well as the two major muscles of the mouth floor, namely the geniohyoid and the mylohyoid muscles, in their whole.

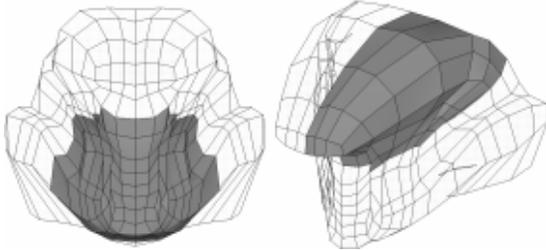

**Figure 2:** Left: modelling of a mouth floor reserction; right: modelling of a left hemiglossectomy

**Motor control of the model**

The tongue model is deformed and controlled by a functional model of muscle force generation mechanisms, namely the Equilibrium Point Hypothesis [4]. Motor commands have been first inferred for the original structure for the three studied vowels, and simulations were then carried out for the various surgery conditions with these original muscles' motor commands hold during 200 ms. Motor commands selection was based on considerations on the tongue shapes in the mid-sagittal plane [5] combined with published EMG data [6][7].

**From tongue shapes to acoustic properties**

The final tongue surface was interpolated by natural cubic spline curves. Then, intersections between the different articulators and a 3D semi-polar grid were computed to estimate the vocal tract area function. The associated formants were finally computed and compared with each others.

### III. RESULTS

*A. Impact of a left hemiglossectomy*

Only the results for the second case (intermediate level of activation for the sectioned fibers) are presented, most fibers being either intact or fully removed after resection.

*(a) Impact on the tongue mobility*

After a hemiglossectomy, we noticed an important deviation of the apex, either on the healthy tissue side for vowels /u/ and /i/ (Fig. 3) or on the flap side for vowel /a/, as well as its rotation. The deviation is more or less important for the different vowels according to the flap biomechanical properties. After reconstruction, the smaller the stiffness of the flap, the larger the asymmetry of the tongue shaping. This is especially true for vowels /i, u/, due to the styloglossus activation, but also for vowel /a/, probably due to the combined activation of the anterior genioglossus and hyoglossus, two muscles slightly effected by the exeresis. In the case of vowel /a/, we also found a more important flattening of the tongue with decreasing flap stiffness: a high stiffness flap restrict the tongue movements.

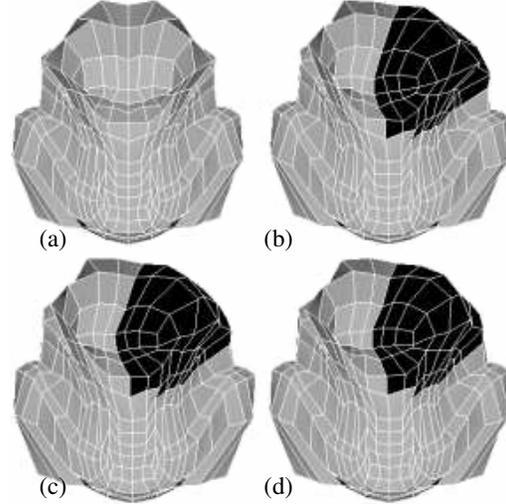

(a)   (b)

(c)   (d)

**Figure 3 :** Impact of a left hemiglossectomy on the tongue symmetry for vowel /i/. (a): non pathological case, (b)-(d): reconstruction with flaps of increasing stiffness (0.2, 1 or 5 times the stiffness of passive tongue tissues).

*(b) Impact on the acoustic signal*

Figure 4 shows the variations of the first two formants associated with the different resections and reconstructions. A left hemiglossectomy (left panel) has a negligible impact on the production of vowels /a/ and /u/. For /i/ the formants deviation is more important, resulting in an average increase of 8% for F1 and average decrease of 9% for F2. In terms of formant changes, a softer flap seems to have less impact, particularly for /a/, but the differences between flaps are slight. These results are coherent with the variations observed on the tongue shapes.

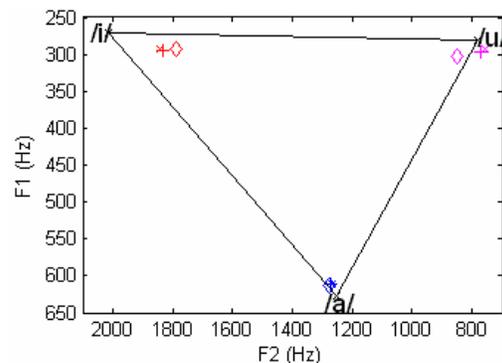

**Figure 4:** F1/F2 formant patterns for a left hemiglossectomy for flaps with different stiffness (small stiffness represented by x-marks, medium

stiffness by crosses and high stiffness by diamonds). Triangles join the extreme vowels obtained with the non-pathological model.

### B. Impact of a large mouth floor resection

The anterior part of the genioglossus being resected, large discrepancies appeared between the three cases studied concerning the sectioned fibers level of activation. The simulations showed that the smaller the activity of the sectioned fiber, the more important the differences with the non pathological case. The implementation leading to a perfectly symmetrical model, no rotation of the tongue was possible.

#### (a) Impact on the tongue mobility

The simulations revealed a large impact of mouth floor resection on tongue elevation and protraction movements, for vowels /u/ and /i/. The mylohyoid muscle allows the rigidification of the mouth floor, essential to its elevation. Furthermore, the posterior genioglossus is the main muscle involved in protraction movement. Its partial resection limits the contraction of the anterior part of the tongue base. For vowels /a/, a high stiffness flap limits the tongue mobility and prevents the limits the flattening of the tongue.

For the different vowels, a high stiffness flap seems the most appropriate voice. Figure 5 shows the results for vowels \i\ for with the different reconstruction schemes and for the non pathological case. A high stiffness flap favors the tongue protraction movements whereas a small stiffness flap can lead to a total obstruction of the vocal tract. Similar results were observed for vowels \u\ and \a\ (reduction of the airway section in the pharyngeal area).

The hypotheses made concerning the activation of the sectioned fibers lead to significant differences: obstruction or not of the vocal tract for vowel \i\, backward rotation of the apex for vowel \a\ in the absence of activation (inactivation of the anterior genioglossus that cannot counteract anymore the activation of the hyoglossus) and backward movement more or less pronounced for \i,a,u\. Comparison of simulation results with data collected on patients could shed light on the hypothesis (no activation, partial activation or full activation) that seems to be the most realistic. However, the choice of the activation did not impact the effect of the flap properties on the tongue mobility.

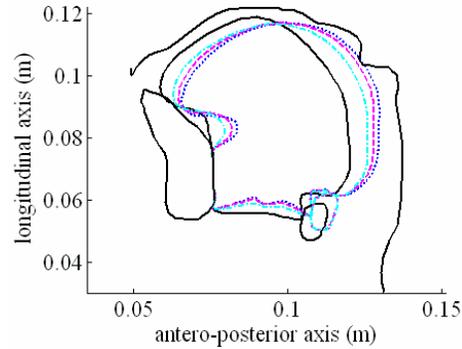

**Figure 5:** Shape of the tongue in the mid-sagittal plane after a mouth floor resection for vowel \i\ (mid level of activation for the sectioned fibers). The plain contour represents the non pathological case, the dotted contour the reconstructed model with the small stiffness flap, the dashed contour the medium stiffness flap and the dashed-dot contour the high stiffness flap.

#### (b) Impact on the acoustic signal

Figure 6 shows plots the first and second formants for vowel \i\ for the partial and full activation hypotheses. Results can be summarized as follows :

- A large mouth floor resection seems to have severe consequences on speech production. For vowel /u/, keeping the motor commands inferred for non- pathological conditions leads to an obstruction of the vocal tract in the pharyngeal region, due to the resection of the anterior part of the posterior genioglossus that counteracted the effects of the styloglossus activation before surgery. Therefore, not formant could be computed.
- The current pattern of activation did not permit to produce the high front vowel /i/ (average increase of 23% for F1 and average decrease of 17% for F2), with important discrepancies according to the flap. A high stiffness flap leads to a higher increase of F1, whereas a small stiffness flap leads to a higher decrease of F2.
- For vowel /a/, we can observe a decrease in F1 and F2, particularly for low stiffness flaps, correspond to a deviation from vowel /a/ to vowel /o/.

Combined with the tongue shapes observation, our results show that for mouth floor resection high stiffness flap should be favoured. Indeed, only this kind of flap can allow the tongue to reach a front high position close to /i/.

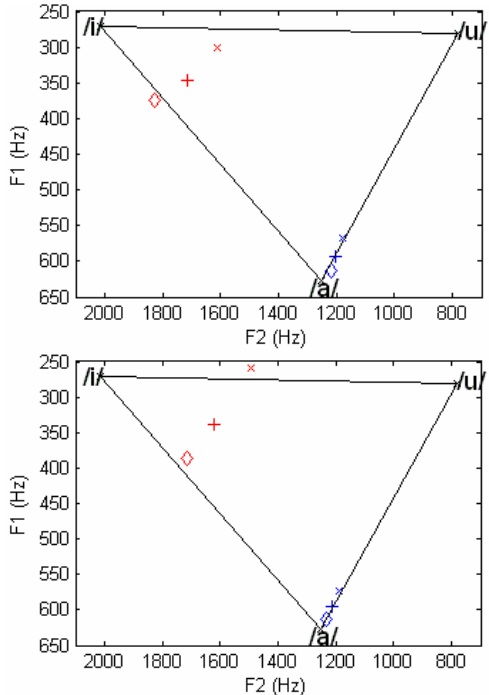

**Figure 6:** F1/F2 formant patterns for a mouth floor resection for flaps with different stiffness (small stiffness represented by x-marks, medium stiffness by crosses and high stiffness by diamonds). Triangles join the extreme vowels obtained with the non-pathological model. Top panel: no activation, bottom panel low activation for the sectioned fibres.

### III. DISCUSSION

Simulations with a realistic 3D biomechanical model could be of a significant improvement in planning tongue surgery systems. In terms of F1/F2 patterns changes our results are in good agreement with measurements made on patients [8]. The role of the flap stiffness on tongue mobility could also be assessed and, interestingly, it is different for the hemiglossectomy than for the mouth floor resection. Further improvements of the model include algorithmic aspects aiming at a significant decrease of the computation time and mesh matching methods to design patient specific oral cavity models.